\documentclass[12pt,preprint]{emulateapj}

\shorttitle{PAH contribution to the IR output energy at z$\simeq$2}
\shortauthors{G. Lagache et al.}

\begin{document}


\title{PAHs contribution to the infrared output energy of the Universe at z$\simeq$2}


\author{G. \ Lagache\altaffilmark{1}, 
H. \ Dole\altaffilmark{1,2}, J.-L. Puget\altaffilmark{1},
P. G. \ P\'erez-Gonz\'alez\altaffilmark{2}, E. Le Floc'h\altaffilmark{2}, 
G. H. Rieke\altaffilmark{2}, C. Papovich\altaffilmark{2}, E. Egami\altaffilmark{2}, 
A. Alonso-Herrero\altaffilmark{2}, C. W. Engelbracht\altaffilmark{2}, 
K.D. Gordon\altaffilmark{2}, K. A. Misselt\altaffilmark{2},  
and J. E. Morrision\altaffilmark{2}}

\altaffiltext{1} {Institut d'Astrophysique Spatiale, b\^at 121,
Universit\'e Paris-Sud, F-91405 Orsay Cedex}
\altaffiltext{2} {Steward Observatory, University of Arizona, 933 N Cherry Ave,
Tucson, AZ 85721, USA}



\begin{abstract}
We present an updated phenomenological galaxy evolution model to fit
the {\it Spitzer} 24, 70 and 160~$\rm \mu$m number counts as well 
as all the previous mid and far infrared observations.
Only a minor change of the co-moving luminosity density 
distribution in the previous model (Lagache, Dole, Puget 2003), combined with a slight modification
of the starburst template spectra mainly between 12 and 30~$\rm \mu$m, 
are required to fit all the data available.
We show that the peak in
the MIPS 24~$\rm \mu$m counts is dominated by galaxies
with redshift between 1 and 2, with a non negligible
contribution from the z $\ge$ 2 galaxies ($\sim$ 30$\%$ at S=0.2 mJy).
The very close agreement between the model and number
counts at 15 and 24~$\rm \mu$m strikingly implies that (1) 
the PAHs (Policyclic Aromatic Hydrocarbons) features remain prominent in the redshift band 0.5 to 2.5
and
(2) the IR energy output has to be dominated
by $\sim$3 10$^{\rm 11}$~L$_{\rm \odot}$ to
$\sim$3 10$^{\rm 12}$~L$_{\rm \odot}$ galaxies from redshift 0.5
to 2.5.
Combining {\it Spitzer} with the {\it Infrared Space Observatory (ISO)}
deep cosmological surveys gives for the first time 
an unbiased view of the infrared Universe from z=0 to z=2.5.
\end{abstract}




%
\section{Introduction}
A Cosmic far-Infrared Background (CIB) that would trace the 
peak of the star--formation and metal production in galaxy assembly 
has long been predicted.
The first observational evidence for such a background was reported by Puget et al.
(1996) (see Hauser \& Dwek 2001, for a review).  
This discovery, together with recent cosmological surveys in 
the infrared (IR) and submillimeter, has opened new perspectives on our 
understanding of galaxy formation and evolution. The surprisingly large 
amount of energy in the CIB shows that it is crucial to 
probe the galaxies contributing to it to understand when and how the bulk 
of stars formed in the Universe.
\\
To understand the sources contributing to the CIB
and to interpret the deep source counts, we have developed a 
phenomenological model that 
constrains in a simple way the IR luminosity function evolution with redshift, 
and fits all the existing source counts consistent with the
redshift distribution, the CIB intensity, and, for the first time, the CIB
fluctuation observations from the mid-IR to the submm range
(Lagache et al. 2003).
{\it Spitzer} has provided deep new multiwavelength source counts 
from 24 to 160 $\rm \mu m$ (Papovich et al. 2004,
Dole et al. 2004a). We use these new data to search for a new
optimisation of the Lagache et al. (2003) model parameters. A remarkable
result is that only a minor change of the co-moving luminosity density 
distribution combined with a slight modification of the starburst
model spectra, mainly for the 12 $\le \lambda \le$30~$\rm \mu$m range,
are required to fit all the previous
data together with the new constraints.
 
%
\section{The model}
\label{sect:model}
\begin{figure}
\plotone{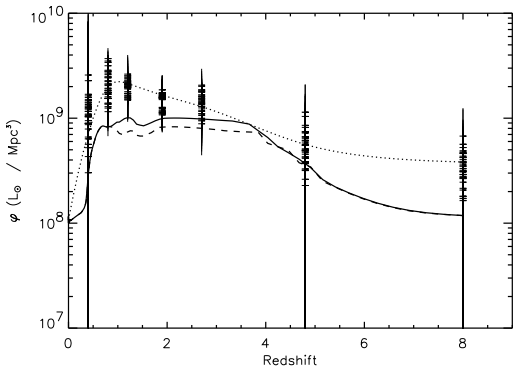}
\caption{\label{phi} Co-moving luminosity density distribution of the Lagache et al.
(2003) model (dashed line) and the updated model presented here
(continuous line).
Also shown for comparison
is the co-moving luminosity density distribution from all
cases of Gispert et al. 2001 (crosses with error bars), 
together with the best fit passing through all cases (dot line).}
\end{figure}

\begin{figure}
\plotone{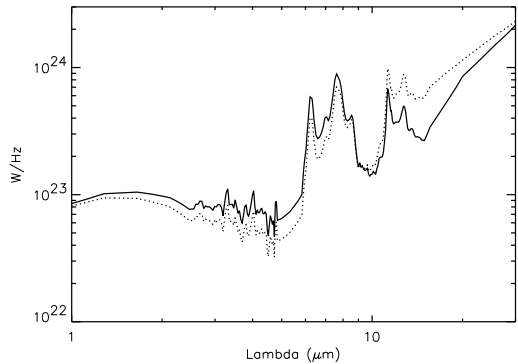}
\caption{\label{spectrum} Starburst model spectrum for L =3 10$^{11}$ L$_{\odot}$.
Dashed line: Lagache et al. (2003), continuous line: the present updated version
of the model.
The spectrum has not been modified for $\lambda \ge$30~$\rm \mu$m. The biggest
change is about a factor 2 (for a 3 10$^{11}$ L$_{\odot}$ galaxy) 
around 15~$\rm \mu$m.}
\end{figure}

The model is discussed extensively in Lagache et al. (2003).
Our goal is to build the simplest model with the fewest free parameters and separate ingredients.
We fix the cosmology to $\Omega_{\Lambda}$=0.7, $\Omega_{0}$=0.3 and h=0.65.
We assume that IR galaxies are 
mostly powered by star formation and hence we use Spectral Energy Distributions (SEDs)
typical of star-forming galaxies\footnote{This assumption is based 
primarily on observations by ISOCAM and ISOPHOT, but is confirmed 
by the first {\it Spitzer} studies of galaxy SEDs in the Lockman Hole and Groth Strip
(Le Floc'h et al. 2004; Alonso Herrero et al. 2004)}. 
Although some of the galaxies will have AGN-dominated SEDs, they are 
a small enough fraction ($<$ 10\%: Alonso Herrero et al.  2004) that
they do not affect the results significantly.
We therefore construct 'normal' and starburst galaxy template SEDs: 
a single form of SED is associated with each activity type and luminosity.
We assume that the Luminosity Function (LF) is represented by these two 
activity types and that they evolve independently. We search for the 
form of evolution that best reproduces the existing data.
The new optimisation of the model parameters
reproduces: (i) the number counts at 15, 24, 60, 70, 90, 160, 170,
 and 850~$\rm \mu$m, (ii) the known redshift distributions (mainly at 15 and 170~$\rm \mu$m),
(iii) the local luminosity functions at 60 and 850~$\rm \mu$m,
(iv) the CIB (from 100 to 1000~$\rm \mu$m) and its fluctuations
(at 60, 100 and 170~$\rm \mu$m).\\
Compared with the form of the model derived by Lagache et al. (2003), 
only a slight change of the co-moving luminosity density 
distribution is required (Fig. \ref{phi}), together with minor modifications
to the starburst template spectra mainly between 12 and 30~$\rm \mu$m (Fig. \ref{spectrum}). 
The modified starburst spectra still reproduce the color diagrams of Soifer \& Neugeubauer 
1991 (12/25 versus 60/100). We do not modify any other parameters
in the model, nor the SED of the 'normal' galaxy template nor
the SED of the starburst galaxy templates at longer wavelengths, nor the parametrisation
of the local LF.\\

PAHs radiate about 10$\%$ of the 1-1000~$\rm \mu$m luminosity in a set of features concentrated
mostly between 6 and 15~$\rm \mu$m (Fig. \ref{spectrum}). 
This set of features is the strongest spectral
concentration known in galaxy spectra. The features affect strongly the
mid-IR number counts when the redshift brings them into the bandpass filter.
For the 15~$\rm \mu$m ISOCAM observations, the peak contribution arises when the 7.7~$\rm \mu$m
maximum is close to the middle of the band i.e. z$\sim$0.8. The ISOCAM
15~$\rm \mu$m counts showed the existence of a maximum
in the redshift distribution just at that redshift.
Futhermore, it leads to a maximum in the counts (normalised to the Euclidian,
e.g. on Fig. \ref{number_counts1}),
which can exist only when the IR energy output in the critical redshift range is dominated
by a well defined luminosity range.
Our model predicted a similar behaviour for the 24~$\rm \mu$m band.

\begin{deluxetable}{llccc}
\tablewidth{0pt}
\tablecaption{1 $\rm \sigma_c$ confusion noise values using the best confusion
estimator of Dole et al. (2003), confusion limit $\rm S_{lim}$ and the value
of $\rm q = S_{lim} / \sigma_c$  \label{tab:conf}}
\tablehead{
\colhead{} &
\colhead{24 $\rm \mu$m} &
\colhead{70 $\rm \mu$m } &
\colhead{160 $\rm \mu$m} 
}
\startdata
1 $\rm \sigma_c$ & 8.0 $\rm \mu$Jy& 0.47 mJy & 10.6 mJy \\ 
$S_{lim}$ & {\bf 56 $\rm \mu$Jy} & {\bf 3.2 mJy} & {\bf 39.8 mJy} \\ 
q & 7.0 & 6.8 & 3.8 \\ 
\enddata
\tablenotetext{a}{Using the source density criterion}
\tablenotetext{b}{In this case, the photometric and source density criteria agree}
\end{deluxetable}

\section{Results}
\label{sect:results}

\subsection{Source counts}
\label{sect:counts}
\begin{figure}
\plotone{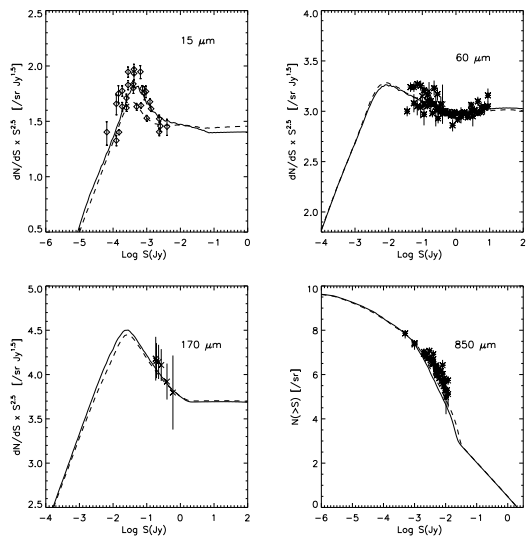}
\caption{\label{number_counts1} Number counts at 15, 60, 170 and 850~$\mu$m (in Log)
together with the model predictions (present work: continuous line, Lagache et
al. 2003: dashed line).
Data at 15~$\rm \mu$m are from Elbaz et al. (1999), at 170~$\rm \mu$m from Dole et al. (2001),
at 60~$\rm \mu$m from Hacking \& Houck (1987), Gregorich et al. (1995), Bertin et al. (1997), 
Lonsdale et al. (1990), Saunders et al. (1990) and Rowan-Robinson et al. (1990) and at 850~$\rm \mu$m 
from 
Smail et al. (1997), Hughes et al. (1998), Barger et al. (1999), Blain et al. (1999), 
Borys et al. (2002), Scott et al. (2002) and Webb et al. (2002).}
\end{figure}

\begin{figure*}
\plotone{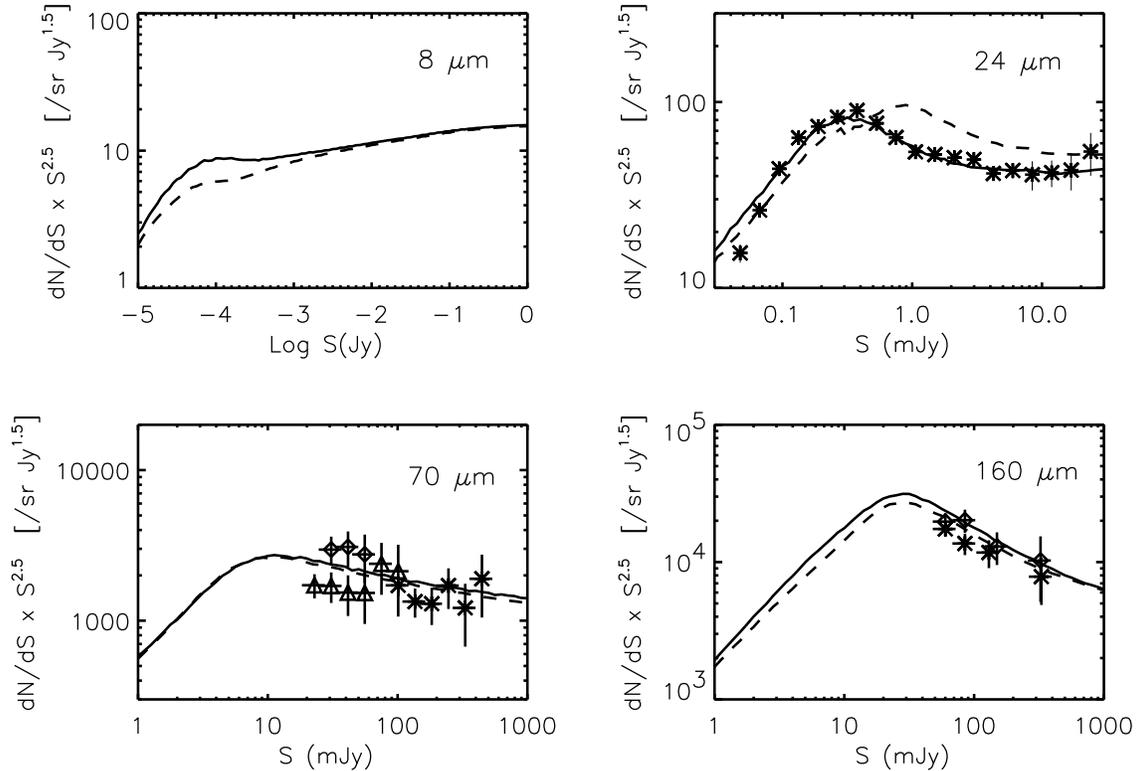}
\caption{\label{number_counts2} Number counts at 24, 70 and 160~$\rm \mu$m 
together with the model predictions (present work: continuous line, Lagache et
al. 2003: dashed line).
Data at 24, 70 and 160~$\rm \mu$m are from Papovich et al. (2004)
and Dole et al. (2004a). Notice that none of the observed source
counts are corrected for uncompleteness.}
\end{figure*}

We show the comparison of the 
number counts at 15, 60, 170 and 850~$\rm \mu$m
with the observations and  as well as the new {\it Spitzer} 24~$\rm \mu$m (Papovich et al. 2004),
70 and 160~$\rm \mu$m (Dole et al. 2004a) counts 
in Fig.~\ref{number_counts1} and Fig.~\ref{number_counts2}.
The updated model reproduces very well the counts at 15 and 24~$\rm \mu$m 
without any significant changes in the other number counts,
as well as the CIB and its fluctuations and the redshift distributions of resolved sources
at 15, 60, 170 and 850 ~$\rm \mu$m. 
The agreement between the observed counts and the model at 24~$\rm \mu$m
is excellent.

\subsection{First interpretation of the {\it Spitzer/MIPS} results}
\label{sect:zdist}

\begin{figure}
\plotone{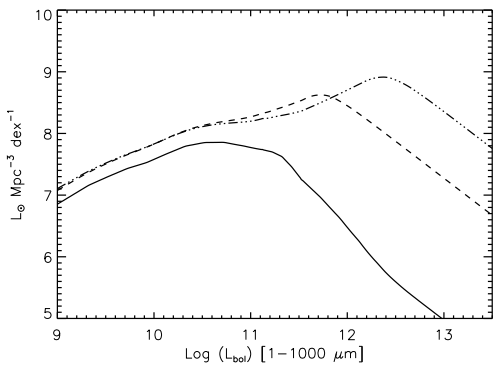}
\caption{\label{LF}  Co-moving evolution of the IR energy output per dex of
luminosity (L dN/dLogL). 
The continuous, dashed, dashed-3dotted 
lines are for z=0, 0.5 and 2 respectively.}
\end{figure}

The consistency
of the model from the mid-IR and far-IR to the submm indicates
that the underlying assumptions are valid up to redshift 2.5.
In particular, the very close agreement between the model and number
counts at 15 and 24~$\rm \mu$m implies that the PAH features
remain prominent in the redshift range 0.5 to 2.5. Furthermore, the well-defined
bump in the 24~$\rm \mu$m number counts (normalised to the Euclidian) at S$\simeq$0.3 mJy
(Fig. \ref{number_counts2}) implies at z$\simeq$2 that L$_{8 \rm \mu m}$= 3.2 10$^{\rm 11}$~L$_{\rm \odot}$
which corresponds to 
L$_{1-1000 \rm \mu m}$=3.5 10$^{\rm 12}$~L$_{\rm \odot}$. These luminosities have
to dominate the IR energy ouput at that redshift (as shown on Fig. \ref{LF}).
As a comparison, an IR energy output dominated by L$_{\star}$ galaxies 
at z$\simeq$2 would maximise the 24~$\rm \mu$m counts around 3~$\rm \mu$Jy
which is totally inconsistent with the observation since most of the CIB
at 24~$\rm \mu$m is resolved at flux densities brighter than $\sim$60~$\rm \mu$Jy.

\begin{figure}
\plotone{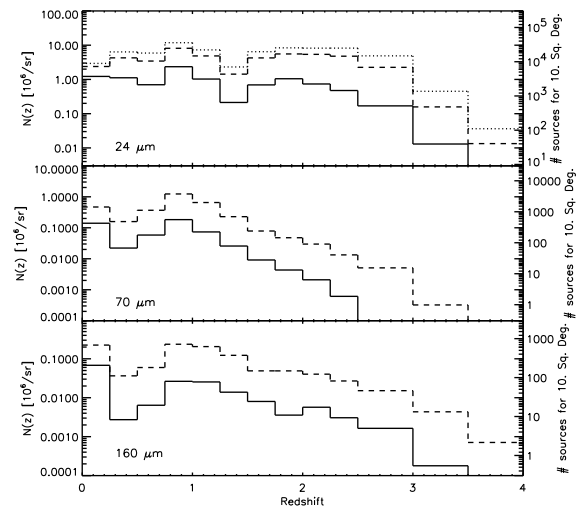}
\caption{\label{zdistrib} MIPS 24, 70 and 160~$\rm \mu$m redshift distributions
(in logarithmic scale).
{\it Solid lines}: Shallow surveys (flux greater than 0.3, 22.3 and 135.1 mJy
at 24, 70 and 160~$\rm \mu$m respectively), {\it dashed line}: Deep surveys
(flux greater than 0.094, 6.8, 50.2 mJy
at 24, 70 and 160~$\rm \mu$m respectively), 
{\it Dotted line}: Ultra Deep 
survey (flux greater than 0.06 mJy at 24~$\rm \mu$m).}
\end{figure}

\begin{figure}
\plotone{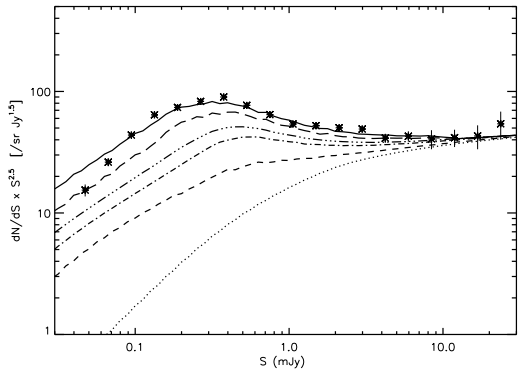}
\caption{\label{24mic} Redshift contribution to the number counts at
24~$\rm \mu$m. The dot, dash, dash-dot, dash-3 dot, long-dash
correspond to the number counts up to redshifts 0.3, 0.8,  1, 1.3 and
2 respectively. Data are from Papovich et al. (2004).}
\end{figure}

Using the new model of
IR galaxy evolution, Dole et al. (2004b) have made new determinations of
the confusion limits for MIPS (Multi Band Imaging Photometer for {\it Spitzer}). 
A summary is provided in Table \ref{tab:conf}. 
Fig. \ref{zdistrib} shows the updated redshift distributions for
such confusion-limited surveys. 
At 24~$\rm \mu$m, the deepest surveys will probe 
star formation up to redshift 3. In the far-IR, MIPS surveys
will probe the largely unexplored window 1$\le$z$\le$2.\\
In Fig~\ref{24mic}, we show the 
different redshift contributions to the 24~$\rm \mu$m number counts.
The peak in the counts is dominated
by galaxies with redshift between 1 and 2. 
There is also a non negligible contribution from $z>$2 galaxies (30$\%$ at S=0.2 mJy).
The integral of the source counts down to 60~$\mu$Jy gives 1.9 nW m$^{\rm -2}$ sr$^{\rm -1}$,
in excellent agreement with Papovich et al. (2004). It represents about
63$\%$ of the CIB at 24~$\rm \mu$m .
It should also be noted from Fig. \ref{24mic} that the redshift distribution changes sharply
for fluxes between 0.3 and 2 mJy. For sources weaker than 150~$\rm \mu$Jy, the
redshift distribution remains about constant.

%
\section{Conclusion}
\label{sect:conclusion}
We have updated the model of Lagache et al. (2003) to 
fit the new constraints from {\it Spitzer} data. 
We show that only a small change of the co-moving luminosity density 
distribution and of the starburst template spectra
are required to fit all 
the observations from the mid-IR to the sub-millimeter.
The agreement between the model and all the data demonstrates that 
the integrated SEDs from galaxies still have prominent
6-9 $\rm \mu$m emission from PAH features up to redshift 2.5. It clearly shows that 
the population of IR galaxies is not dominated by galaxies with featureless continuum spectra
(such as AGN-type SEDs in the mid-IR). The fraction of the energy radiated
by the PAHs is about one tenth of the total IR radiation
locally and remains roughly constant between redshifts 0.5 and 2.5.
\\
Our initial model predicted that the IR output energy had to be
dominated by galaxies with luminosities increasing rapidly with
redshift (L$_{1-1000 \rm \mu m}$$\simeq$ 3 10$^{\rm 12}$~L$_{\rm \odot}$
at z=2.5). This was needed mainly to account for the SCUBA counts
(the total energy ouput at that redshift being mostly driven by the submillimeter
extragalactic background level). The excellent agreement between the 24~$\rm \mu$m
counts and the model confirms strikingly the evolution of the IR luminosity function
as being dominated in energy by $\sim$3 10$^{\rm 11}$~L$_{\rm \odot}$ to
$\sim$3 10$^{\rm 12}$~L$_{\rm \odot}$  galaxies from redshift 0.5
to 2.5.
\\
We have for the first time with {\it Spitzer} an unbiased
view of the IR Universe up to redshift 2.5.
With all the mid-IR to sub-mm data, the model is now very well
constrained up to z$\simeq$2.5.
Other predictions, together with the number counts of all the planned
and future mid-IR to mm surveys, will be available
on the web page:\\
 \url{http://www.ias.fr/PPERSO/glagache/act/gal$\_$model.html}.\\
\acknowledgments
This work is based on observations made with the {\it Spitzer} Space
Telescope, which is operated by the Jet Propulsion Laboratory,
California Institute of Technology under NASA contract 1407.
We thank the funding from the MIPS project, which is supported by
NASA through the Jet Propulsion Laboratory, subcontract \#960785.
GL warmly thanks the MIPS IT in Tucson, HD and JLP for this fruitful collaboration
and G.H.R for his careful reading of the manuscript.


\begin{thebibliography}{} 
\bibitem{Alonso} Alonso Herrero et al., 2004, this issue
\bibitem{Barger} Barger A.J., Cowie L.L., Sanders D.B., 1999, ApJ 518, L5
\bibitem{Bertin} Bertin E., Dennefeld M., Moshir M., 1997, A\&A 323, 685
\bibitem{Blain} Blain A.W., Kneib J.-P., Ivison R.J. et al., 1999, ApJ 512, L87
\bibitem{Borys} Borys C., Chapman S., Halpern M. et al., 2003, MNRAS, 344, 385
\bibitem{Dole01} Dole H., Gispert R., Lagache G. et al., 2001, A\&A 372, 364
\bibitem{Dole03} Dole H., Lagache G., Puget J.-L., 2003, ApJ, 585, 617
\bibitem{Dolea} Dole H., Le Floc'h E., Papovich C. et al., 2004a, ApJ, this issue
\bibitem{Doleb} Dole H., Rieke G.H., Lagache G. et al., 2004b, ApJ, this issue
\bibitem{Elbaz} Elbaz D., Cesarsky C.J., Fadda D. et al., 1999, A\&A 351, 37
\bibitem{Gispert} Gispert R., Lagache G., Puget J.-L., 2000, A\&A 360, 1
\bibitem{Gregorich} Gregorich D.T., Neugebauer G., Soifer B.T., Gunn J.E., Herter T.L., 1995, AJ 110, 259
\bibitem{Hauser} Hauser M.G., Dwek E., 2001, ARAA, 37, 249
\bibitem{Hacking} Hacking P.B., Houck J.R., 1987, ApJS 63, 311
\bibitem{Hughes} Hughes D.H., Dunlop J.S., Rowan-Robinson M. et al., 1998, Nature 394, 241
\bibitem{Lagache} Lagache G., Dole H., Puget J.-L., 2003, MNRAS 338, 555
\bibitem{Lefloch} Le Floc'h E., et al., 2004, ApJ, this issue
\bibitem{Lonsdale} Lonsdale C.J., Hacking P.B., Conrow T.B., Rowan-Robinson M., 1990, ApJ 358, 20
\bibitem{Puget} Puget J.-L., Abergel A., Bernard J.-P. et al., 1996, A\&A 308, L5 
\bibitem{Papovich} Papovich C., Dole H., Egami E., 2004, ApJ, this issue
\bibitem{Rowan} Rowan-Robinson M., Saunders W., Lawrence A., Leech K., 1991, MNRAS 253, 485
\bibitem{Saunders} Saunders W., Rowan-Robinson M., Lawrence A. et al. 1990, MNRAS 242, 318
\bibitem{Scott} Scott S.E., Fox M.J., Dunlop J.S. et al., 2002, MNRAS 331, 817
\bibitem{soifer91} Soifer B.T., Neugebauer G., 1991, AJ 101, 354
\bibitem{Smailb} Smail I., Ivison R.J., Blain A.W., 1997, ApJ 490, L5
\bibitem{Webb} Webb T.M.A., Eales S.A., Lilly S.J. et al., 2003, ApJ, 587, 41
\end{thebibliography}
\end{document}